\documentclass[11pt,twoside]{article}
\usepackage{asp2010}
\usepackage{graphicx}

\resetcounters
\allowtitlefootnote

\begin{document}

\title{{\sc AMBER} high-spectral resolution interferometry of Be stars:
       More than just stars and disks?
       }
\author{S.~\v{S}tefl$^1$, T.~Rivinius$^1$, D.~Baade$^{2}$, A. Carciofi$^3$
\affil{$^1$ European Organisation for Astronomical Research in the Southern Hemisphere,
      Casilla 19001, Santiago 19, Chile}
\affil{$^2$European Organisation for Astronomical Research in the
      Southern Hemisphere, Karl-Schwarzschild-Str.~2, 85748 
      Garching bei M\"{u}nchen, Germany}
\affil{$^3$Instituto de Astronomia, Geof{\'\i}sica e Ci{\^e}ncias
      Atmosf{\'e}ricas, Universidade de S\~ao Paulo, Rua do Mat\~ao 1226,
      S\~ao Paulo, SP 05508-900, Brazil}}
\titlefootnote{Based on observations collected 
at the European Southern Observatory, Chile (Prop.\ Nos. 282.D-5014, 
082.D-0146, 383.D-0522, 284.D-5043, 084.D-0355, 084.C-0848, 385.D-0642, 
60.A-9053.}

\begin{abstract}
High spectral-resolution Br\,$\gamma$ VLTI/{\sc AMBER} observations of classical Be stars reveal 
complex visibility, phase, and closure phase profiles, which were not seen in previous 
lower-resolution data.  They present new challenges for the modeling of Be stars and may 
require additional ingredients over and above the conventional central star + circumstellar 
disk paradigm.
\end{abstract}

\section{Motivation and targets}
{\sc AMBER} \citep{2007A&A...464....1P}
was the first high-spectral resolution (HR, R~$=$~12\,000) interferometric instrument and still 
is the only one capable of taking closure phase data in this mode.  After the commissioning in 
the end of 2008, the spectral resolution was increased eightfold wrt. its medium-resolution mode, 
and it is about 30 times as high as  that of CHARA/{\sc MIRC} \citep{2004SPIE.5491.1370M}. The present 
(2011) list of standard HR settings comprises 14 central wavelengths but most of the actual HR 
observations cover the Br~$\gamma$ spectral region. 

The order-of-magnitude increase in spectral resolution entails the potential for a qualitative 
leap forward in studies of objects with a considerable range in dynamics in general and rotationally 
supported disks of young or evolved stars in particular, which account for a large fraction of all 
available VLTI observations.  

In order to explore this potential for classical Be stars, we have combined data from three 
{\sc AMBER} projects of our own with selected archival observations.  The objects in 
Tab.~\ref{targets} include almost all Be stars observed to date with {\sc AMBER} in its HR mode. 
Only Achernar ($\alpha$\,Eri) was not included because it had but a tiny disk at the time of the 
{\sc AMBER} observations. Observations were done both with the 8.2m diameter Unit Telescopes (UTs) 
and the 1.8m Auxiliary Telescopes (ATs). Because even the largest circumstellar Be disk diameters in 
Br\,$\gamma$ are of a few miliarcseconds, the longest baselines -- UT1-UT3-UT4 or UT1-UT2-UT4 triplets, 
D0-H0-G1-I1 or A0-K0-G1-I1 ATs quadruplets -- were used for observations. All data were reduced with
the help of the yorick amdlib3 reduction package \citep{2007A&A...464...29T,2009A&A...502..705C}.

\begin{table}[!ht]
\caption{Classical Be stars observed in {\sc AMBER} HR mode and Br\,$\gamma$ 
region in 2008-2010. Parameters of the targets are taken from the Simbad database.}
\smallskip
\begin{center}
{\small
\begin{tabular}{rrlrrrl}
\tableline
\noalign{\smallskip}
Star             &  HD    & Spectral & v $\sin i$ & $K_{\mathrm{mag}}$   & parallax  & Epoch          \\
                 &        &  type    & [km/s]     & [mag]             &  [mas]    &                \\
\noalign{\smallskip}     
\tableline
\noalign{\smallskip}
$\zeta$\,Tau     &  37202 & B2IV     &  310       &         2.81      &  7.82     &  Oct 15, 2008  \\
$\alpha$\,Col    &  37795 & B7IVe    &  195       &         2.83      & 12.16     &  Jan 9,10, 20, 2010  \\  
$\kappa$\,CMa    &  50013 & B1.5IVe  &  210       &         3.55      &  4.13     &  Jan 9, 10, 17, 2010   \\
$\omega$\,CMa    &  56139 & B2IV-Ve  &  105       &         4.37      &  3.53     &  2008-2010, large dataset  \\
$\beta$\,CMi     &  58715 & B8Ve     &  210       &         3.10      & 19.16     &  Dec 31, 2009  \\ 
$\omega$\,Car    &  89080 & B8IIIe   &  220       &         3.45      &  8.81     &  Dec 21, 2008 \\
48\,Lib          & 142983 & B5IIIp(?)&  395       &         4.59      &  6.36     &  May 12, 2009 \\
$\delta$\,Sco    & 143275 & B0.2IVe  &  165       &         2.43      &  8.12     &  2009-2011; continuing obs. \\
\noalign{\smallskip}
\tableline
\label{targets}
\end{tabular}
}
\end{center}
\end{table}

\begin{figure}[!ht]
\includegraphics[width=\textwidth,clip=true]{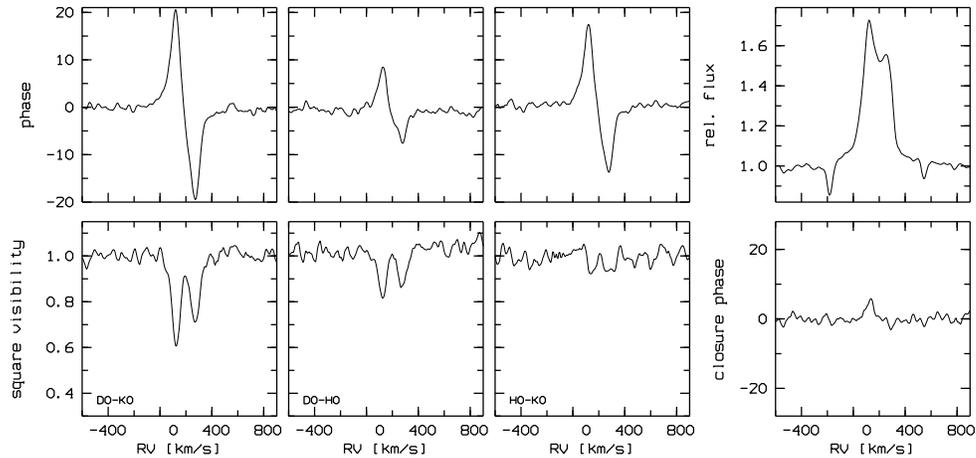}
\caption{$\kappa$\,CMa observed on January 9, 2010, AT baseline configuration D0-H0-K0. The baseline lengths 
    and position angles are: D0-H0 64.0m@71.0$^{\circ}$, H0-K0 32.0m@71.0$^{\circ}$ and D0-K0 96.0m@71.0$^{\circ}$. Although
    canonical S-shaped profiles appear in phase diagrams,  a double-peak reversal can be seen in all 
    visibilities.}
\label{kCMaint}
\end{figure}

\section{Results}

The generally expected -- and at low and medium spectral resolution often confirmed -- 
scheme of the interferometric appearance of circumstellar disks results from simple uniform 
or Gaussian disk models and consists of a single central reversal (akin to a spectral flux 
profile with a stellar absorption line) in the visibility and a smooth S-shaped phase profile.  
The canonical phase profile as exhibited by $\kappa$\,CMa is shown in Fig.~\ref{kCMaint}.

\begin{figure}[!ht]
\includegraphics[width=\textwidth,clip=true]{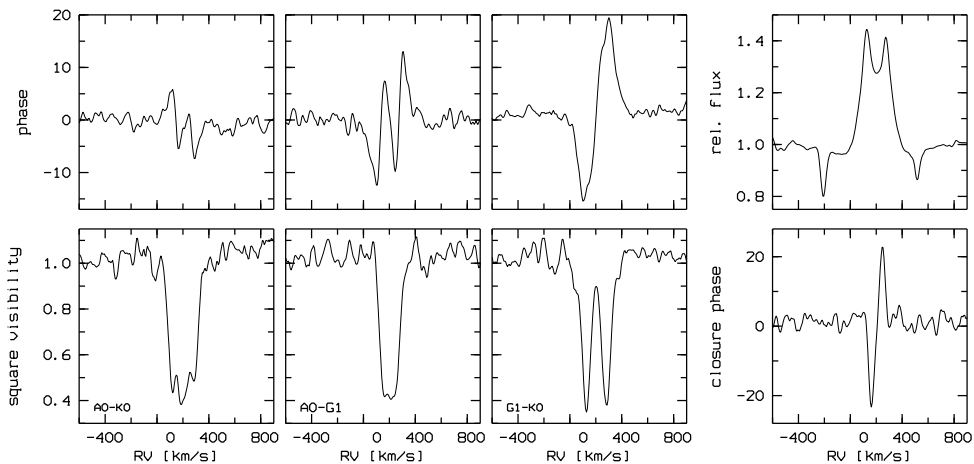}
\caption{$\alpha$\,Col observed on January 20, 2010 (HJD\,55216.083), AT telescopes
     at the baseline configuration A0-K0-G1. The baseline lengths and position angles 
     are: A0-K0 128.0m@71.0$^{\circ}$, K0-G1 90.5m@26.0$^{\circ}$ and A0-G1 90.5m@116.0$^{\circ}$.}
\label{aColint}
\end{figure}

\begin{figure}[!ht]
\includegraphics[width=\textwidth,clip=true]{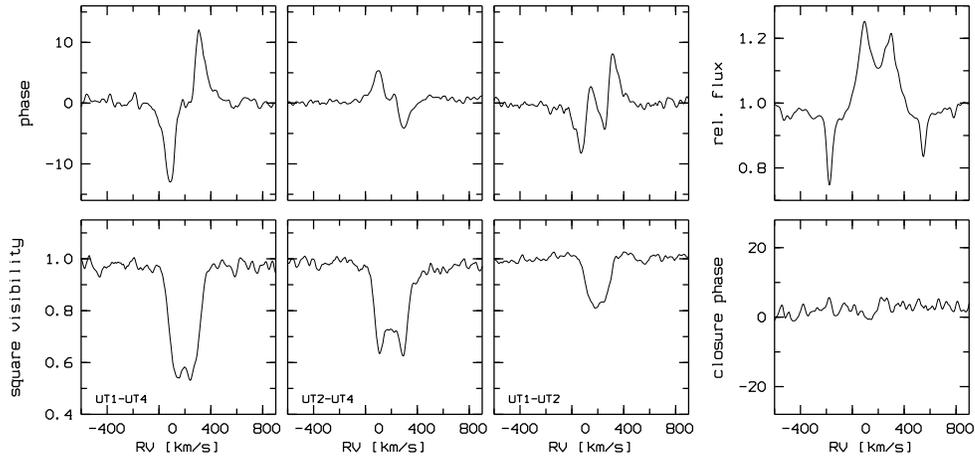}
\caption{$\beta$ CMi observed on December 31, 2009 (HJD 55196.190), baseline configuration UT1-UT2-UT4. 
    The baseline lengths and position angles are: UT1-UT2 56.6m@26.2$^{\circ}$, UT2-UT4 89.4m@81.3$^{\circ}$ 
    and UT1-UT4 130.2m@60.4$^{\circ}$.}
\label{bCMiint}
\end{figure}

By contrast, HR {\sc AMBER} observations of Br-$\gamma$ square visibilities, phases, spectroscopic 
normalized fluxes, and closure phases in all Be stars in Tab.~\ref{targets} reveal much 
more complex structure (Figs.~\ref{kCMaint}, \ref{aColint} and \ref{bCMiint}).  Broadened, double- 
or multiple-peak 
profiles commonly appear in the visibilities. In phase, a third peak is sometimes seen superimposed 
to the center of the expected S-shaped profile.  In some cases, this phase reversal merely is 
a small perturbation of the S-shaped profile, but in others it reaches an amplitude comparable to 
the one of the main peaks.  Both visibility and phase profiles are very strongly dependent on the 
projected position angle of the baseline used for the observations.  Similar patterns are not known 
from interferometry of targets other than Be stars and so are asking for a special interpretative 
effort.

\noindent
The following is an attempt to briefly describe the most prominent features in the visibility and 
phase profiles of the stars in the sample.  Remember that their appearance is strongly dependent 
on the projected position angle of the baseline.  
\begin{description}
\item[$\zeta$\,Tau:] The only HR observation was obtained 
    during the {\sc AMBER} HR science verification run in October 2008 and is of relatively low quality.  
    An asymmetric double peak visibility profile emerges at all baselines. A sharp peak of 
    different height is superimposed to a shallow S-shape phase profile.  
\item[$\alpha$\,Col:] Visibilities possess a double- or multi-peak structure, and  phases show 
    a strong reversal at two baselines. The closure phase  indicates a very strong disk asymmetry, 
    see Fig.~\ref{aColint}.
\item[$\kappa$\,CMa:] A double-peak structure in the visibility is seen even at shorter baselines, 
    S-shaped phase profiles at all baselines.
\item[$\beta$\,CMi:]  A phase reversal and a double/multiple visibility structure appear at all 
    baselines for this very nearby Be star although the closure phase only suggests a negligible 
    disk asymmetry, see Fig.~\ref{bCMiint}.
\item[28\,($\omega$)\,CMa:] A pole-on view of the star is offered from H\,$\alpha$ emission line profiles 
     and was also confirmed by modeling of rapid variations of photospheric lines due to non-radial 
     pulsation. \citet{2003A&A...411..181M} derived the inclination of the rotational axis $i$~=~15$^{\circ}$
     or 24$^{\circ}$. Double-peak visibility profiles in 28\,CMa show that the feature 
     can appear even in stars with supposedly little line-of-sight motions in the disk. At some epochs, 
     a weak phase reversal is present at the G1-K0 baseline, but pure S-shaped profiles result from 
     observations at the other baselines. Analysis of a large dataset obtained in 2009-2010 is in 
     progress (\v{S}tefl et al., in preparation).
\item[48\,Lib:] The huge almost over-resolved disk shows strong double-peak visibility and closure 
     phase profiles, a strong phase reversal appears at one baseline. For a more detailed 
     discussion see \v{S}tefl et al. (A\&A, to be submitted).
\item[$\delta$\,Sco:] Phase reversal and double-peak visibility profiles dominate at some baselines. 
     A comprehensive analysis of all data is in progress (\v{S}tefl et al., in preparation).
\end{description}

\begin{figure}[!ht]
\includegraphics[width=\textwidth,clip=true]{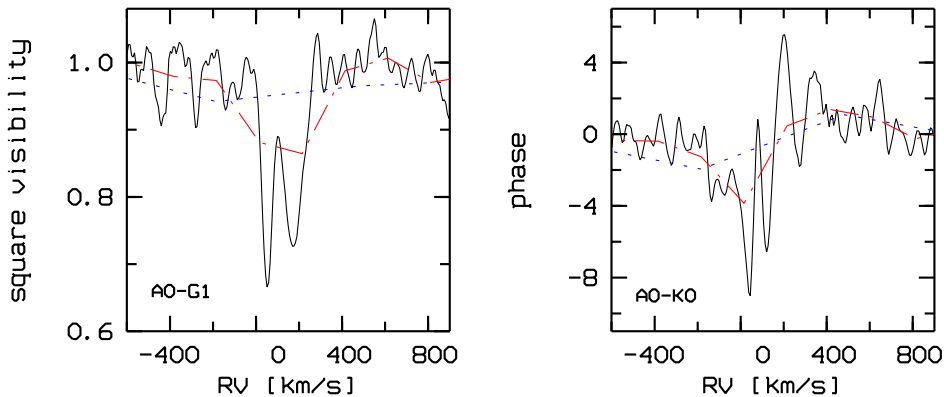}
\caption{The effect of the spectral resolution demonstrated on observations of $\delta$\,Sco obtained on 
    May 10, 2010. Visibility at the A0-G1 baseline and phases at A0-K0 are shown. The full black line 
    represents {\sc AMBER} HR observations, which are rebinned to {\sc AMBER} medium resolution of R=1500 
    (overplotted as the dashed red line) and the highest {\sc MIRC} resolution R=450 (dotted blue line).
    }
\label{resolution}
\end{figure}

The key role of the high spectral resolution for interferometry of Be stars is demonstrated in a simple 
way in Fig.~\ref{resolution}. The double peak visibility at the A0-G1 baseline and phase profile at 
the A0-K0 baseline, showing a strong phase reversal, are rebinned to the {\sc AMBER} medium and 
CHARA/{\sc MIRC} highest resolution.  
The fine structure both in visibility and phase is lost already in the {\sc AMBER} medium resolution
and these characteristics appear as single reversals of reduced amplitude. Its position at the 
{\sc MIRC} resolution is moreover biased by the starting wavelength of the wavelength/RV grid.

\section{Summary and conclusions}

The {\sc AMBER} HR database of classical Be stars documents that both double/multiple visibility 
profiles and phase reversals are common features.  Double/multiple-peak visibility profiles appeared 
in all observed Be stars at some baseline(s) and time(s).  Their occurrence and peak separation do 
not seem to depend on the length of the baseline.  Asymmetric visibility profiles may correspond to 
a transition between single and double peak profiles.  Because not even in the most nearby objects 
in our sample the central stars are resolved, the double-peak visibility profiles are not caused by 
the superposition of stellar and disk visibilities.  
  
Phase reversals occurred in five of the eight observed Be stars: ($\zeta$\,Tau, $\alpha$\,Col, 
$\beta$\,CMi,  48\,Lib, $\delta$\,Sco, and $\omega$\,CMa). The preferential detection of phase 
reversals at long baselines may locate the responsible physical process close to the star. 
In $\beta$\,CMi, the feature is strong also at the 45m UT1-UT2 baseline; however, this is a very 
nearby star.   

Two of the stars showing a phase reversal are Be shell stars seen equator-on ($\zeta$\,Tau, 48\,Lib).  
However, the limited database (only one  HR observation for 4 targets, 3 observations of different quality
for 2 targets) does not permit phase reversals and visibilities with more than one peak to be linked to 
disk position angle. For 28\,CMa and $\delta$\,Sco this point will be discussed in dedicated papers. 
Nevertheless, 
the small sample indicates that these features occur over a broad range of spectral B sub-types and 
disk viewing angles, from equator-on to at least medium inclination.  Their presence in steady-state 
phases of the disks around three stars ($\alpha$\,Col, $\beta$\,CMi, $\kappa$\,CMa) and also shortly 
after an outburst in $\omega$\,CMa suggests that the observed phase reversals are unrelated to a 
special disk evolution phase.   

The implied commonality in Be star disks of multiple-peak visibility profiles and phase reversals 
represents a major challenge for disk models. They do not match the S-shaped disk phase profiles, 
which are observed at lower spectral resolution and form the basis of commonly accepted models.  
Without quantitative modeling, their interpretation is rather speculative.  

Our preliminary efforts (Carciofi et al., in preparation) suggest that secondary dynamical disk effects 
may explain double-peak visibility profiles and phase reversals at baselines perpendicular to the 
assumed equatorial plane (48\,Lib, $\delta$\,Sco). This is crudely consistent with the observed 
position angle-dependence in some stars.  However, the model is struggling with phase reversals at 
baselines parallel to the equatorial plane (as in $\alpha$\,Col or $\beta$\,CMi).  Therefore, an 
alternative view is developed in Rivinius et al.\ (A\&A, to be submitted).  With polar mass outflows, 
it goes beyond the canonical central star $+$ circumstellar disk paradigm for classical Be stars 
but may not be applicable to all classical Be stars.  

Obviously, high spectral resolution adds a new and essential dimension to interferometry of objects 
with significant internal motions.  This is demonstrated by the here presented observations of 
classical Be stars. 

\acknowledgements 
This research has made use of the  \texttt{AMBER data reduction package} of the
Jean-Marie Mariotti Center\footnote{Available at http://www.jmmc.fr/amberdrs} and
the SIMBAD database, operated at CDS, Strasbourg, France.

\bibliography{SStefl.bib}

\end{document}